\documentclass{ptapap}
\author{D.\,Baade}[ESO-DE]
\author{Th.\,Rivinius}[ESO-CL]
\author{A.\,Pigulski}[Wroclaw]
\author{D.\,Panoglou}[Rio]
\author{A.\,Carciofi}[SaoP]
\author{G.\,Handler}[Warsaw]
\author{R.\,Kuschnig}[Wien, Graz]
\author{Ch.\,Martayan}[ESO-CL]
\author{A.\,Mehner}[ESO-CL]
\author{A.F.J.\,Moffat}[Montreal]
\author{H.\,Pablo}[AAVSO]
\author{A.\,Popowicz}[Gliwice]
\author{S.M.\,Rucinski}[Toronto]
\author{G.A.\,Wade}[Kingston]
\author{W.W.\,Weiss}[Wien]
\author{K.\,Zwintz}[Inns]
\affil[ESO-DE]{European Organisation for Astronomical Research in the 
Southern Hemisphere, Karl-Schwarzschild-Str.\ 2, 85748 Garching b.\ M\"unchen,
Germany}
\affil[ESO-CL]{European Organisation for Astronomical Research in the 
Southern Hemisphere, Casilla 19001, Santiago 19, Chile}
\affil[Wroclaw]{Astronomical Institute, Wroc{\l}aw University, Kopernika 11, 
51-622 Wroc{\l}aw, Poland}
\affil[Rio]{Observat\'orio Nacional, Rua General Jos\'e Cristino 77, S\~ao 
Crist\'ov\~ao RJ-20921-400, Rio de Janeiro, Brazil}
\affil[SaoP]{Instituto de Astronomia, Geof{\' i}sica e Ci{\^ e}ncias Atmosf{\' e}ricas,
Universidade de S{\~ a}o Paulo, Rua do Mat{\~ a}o 1226, 
Cidade Universit{\' a}ria, 05508-900 S{\~ a}o Paulo, SP, Brazil}
\affil[Montreal]{D{\' e}partement de physique and Centre de Recherche en Astrophysique du 
Qu{\' e}bec (CRAQ), Universit{\' e} de Montr{\' e}al, C.P.\ 6128, 
Succ.\,Centre-Ville, Montr{\' e}al, Qu{\' e}bec, H3C 3J7, Canada}
\affil[AAVSO]{American Association of Variable Star Observers, 49 Bay 
State Road, Cambridge, MA 02138, USA}
\affil[Wien]{University of Vienna, Institute for Astrophysics, Tuerkenschanzstrasse 17, 1180 Vienna, Austria}
\affil[Graz]{Graz University of Technology, Institute of Communication Networks
and Satellite Communications, Inffeldgasse 12, 8010 Graz, Austria}
\affil[Toronto]{Department of Astronomy \& Astrophysics, University of Toronto, 
50 St.\,George St, Toronto, Ontario, M5S 3H4, Canada}
\affil[Kingston]{Department of Physics, Royal Military College of Canada, PO
Box 17000, Stn Forces, Kingston, Ontario K7K 7B4, Canada}
\affil[Gliwice]{Institute of Automatic Control, Silesian University of Technology, 
Gliwice, Poland}
\affil[Warsaw]{Nicolaus Copernicus Astronomical Center, ul.\,Bartycka 18, 00-716 
Warsaw, Poland}
\affil[Inns]{Universit\"at Innsbruck, Institute for Astro- and Particle Physics, \\
  Technikerstrasse 25/8, A-6020 Innsbruck, Austria}

\title{BRITEning up the Be Phenomenon\footnote{Based on data collected
    by the BRITE Constellation satellite mission, designed, built,
    launched, operated and supported by the Austrian Research
    Promotion Agency (FFG), the University of Vienna, the Technical
    University of Graz, the Canadian Space Agency (CSA), the
    University of Toronto Institute for Aerospace Studies (UTIAS), the
    Foundation for Polish Science \& Technology (FNiTP MNiSW), and
    National Science Centre (NCN).}}

\begin{document}

\maketitle

\begin{abstract}

  {\bf Abstract:} Observations of 25\,Ori much expand the picture
  derived of other early-type Be stars with BRITE and SMEI.  Two
  instead of one difference frequencies rule the variability: (a) The
  lower one, 0.0129\,c/d, is the frequency of events with full
  amplitudes of 100-200\,mmag which may signal mass loss possibly
  driven by the higher one, 0.1777\,c/d.  (b) Much of the entire power
  spectrum is a tightly woven network of combination frequencies: (i)
  Below 0.25\,c/d, numerous frequencies are difference frequencies.
  (ii) Many frequencies above 2.5\,c/d can be represented as sum
  frequencies and in a few cases as harmonics.  (iii) Many frequencies 
  between 1.1 and 1.75\,c/d can be portrayed as parents of
  combination frequencies.  The number and fraction of combination
  frequencies increases steeply with decreasing amplitude and
  and accuracy of the frequency matching. 

\end{abstract}

\section{Introduction}
\label{introduction}
Be stars are extremely rapidly rotating, nonradially pulsating stars
\citep{2013A&ARv..21...69R} surrounded by a Keplerian disk that is
governed by viscous decretion of gas from the central star
\citep{1991MNRAS.250..432L, 2012ApJ...744L..15C}.  The feeder process
of the decretion is often eventlike and probably driven by interacting
nonradial pulsation (NRP) modes \citep[][Baade et al., subm.\ to
A\&A]{1998ASPC..135..343R, 2016A&A...588A..56B}.  Ejecta produced by
such events intercept light from the central star and process and
re-emit it.  Except for viewing angles close to the plane of the disk,
the photometric signatures of such outbursts are brightenings
\citep{2012ApJ...756..156H}.

The aim of the BRITE \citep{2014PASP..126..573W} programme on Be stars
is to search for links between nonradial pulsations and mass loss.  So
far, in $\eta$\,Cen \citep{ 2016A&A...588A..56B} as well as 28\,Cyg
(Baade et al., subm.\ to A\&A), a pair of $g$ modes was found to
couple to form large-amplitude variations with their difference
frequencies.  In $\eta$\,Cen, the difference frequency seems to
continually modulate, or even drive, mass loss.  In addition, one
brightening was seen in 28\,Cyg in more than 150 days of monitoring in
each of two different seasons.  The brightening was strongly modulated
with the 0.052\,c/d difference frequency the amplitude of which was
temporarily increased by a factor of a few.  There was no obvious
major phase discontinuity during the transition from quiescence to the
outburst.  Complementing this description with the variability of the
optical emission lines during cyclicly repeating outbursts of
$\mu$\,Cen \citep[][Sect.\,\ref{discussion}
below]{1998A&A...333..125R} proves that the combination of two (or
more) NRP modes is at the core of the mass-loss process from Be stars.

These outbursts may be compared to the functioning of a valve.  The
long-term variability of emission lines
\citep[e.g.,][]{2013A&A...560A..30Z} and of the continuum flux from
their disks \citep{2002AJ....124.2039K, 2003A&A...401..271C}
demonstrates that the opening of the valve can be much reduced or even
suspended for years because, without mass supply, the disks decay
\citep{2017MNRAS.464.3071V}.  Observations of pulsations have not so
far provided insights into the nature of these long timescales.
Accordingly, at least four different clocking levels may need to be
distinguished:
\begin{list}{--}{\itemsep=0mm\parsep=0mm\topsep=0mm}
\item[1.)]
  $g$-mode frequencies
\item[2.)]
  Difference frequencies between $g$-mode frequencies.  When the valve
  is open, they seem to provide much of the pumping.
\item[3.)]
  A slower clock that opens and closes the valve.
\item[4.)]
  A very slow clock that inhibits the opening of the valve for long times.
\end{list}
It is still unknown whether the hypothetical level 3 and 4 clocks are
different or show any regularity.  If pulsations are involved at all
levels, there should be some reduction gear that can produce
frequencies of $\leq10^{-3}$\,c/d from single $\sim1$\,c/d modes.

All space photometers that have observed Be stars (MOST, SMEI, CoRoT,
{\it Kepler}, and BRITE) found that the frequencies of Be stars
cluster in groups.  Typical group boundaries are 0.0-0.35\,c/d (group
$g_0$), 1.4-1.9\,c/d ($g_1$), and 2.5-3.7\,c/d ($g_2$); all limits
have correlated full ranges of at least $\pm$30\%.  All groups can
have pronounced substructures, even broad gaps, and the ``voids''
between them are not totally empty.  Based on {\it Kepler}
observations of the B8e star KIC\,11971405,
\citet{2015MNRAS.450.3015K} conjectured that $g_1$ mainly comprises
$g$ modes and that most/all of the frequencies in $g_0$ and $g_2$ may
be combination frequencies.  From 15 possible $g$-mode frequencies
they succeeded in constructing five sum frequencies one of which is
the sum of two difference frequencies.  This scheme is in agreement
with the overall architecture of the power spectra of Be stars: Groups
$g_0$ and $g_1$ typically have roughly equal widths whereas $g_2$ is
about twice as wide.  In fact, the scheme was already de facto
discovered with many dozens of combination frequencies in CoRoT
observations of the B0.5\,IVe star HD\,49330 by
\citet{2009A&A...506...95H} but not explicitly commented upon.

\section{BRITE and SMEI observations of 25\,Ori}
\label{observations}
25\,Ori is an early-type Be star, which in global physical properties
as well as viewing perspective closely resembles $\eta$\,Cen and
28\,Cyg.  In 2014/15, its overall light curve was governed by two
modes of variability: (i) the difference frequency of 0.1777\,c/d
between a pair of $g$ modes and (ii) two brightenings during which the
amplitude associated with 0.1777\,c/d was boosted by nearly an order
of magnitude to full ranges of 100-200\,mmag whereas the amplitudes of
its strongest parent modes only grew by a factor of 2-3.  The improved
frequency resolution of the combination of the BRITE data with SMEI
\citep{2004SoPh..225..177J} observations between 2003 and 2010 shows
that 0.1777\,c/d is the difference frequency of several pairs of
frequencies most of which are in the range of plausible $g$ modes.
Like in 28 Cyg, there were no major phase jumps of the difference
frequency during the transitions between quiescence and brightenings.
Accordingly, the two events in 25\,Ori appeared to be phase coherent
w.r.t.\ the 0.1777\,c/d variability.  They followed each other within
about 78\,d.

In the SMEI data, a frequency of 0.0129\,c/d or period of 77.5\,d
exists so that the brightenings in 2014/15, and putative mass-loss
events, in 25\,Ori seem to repeat with this period.  At 18\,mmag, the
nominal semi-amplitude exceeds all others by almost a factor of two;
however, the 100-200\,mmag events are part of this activity.  A slow
decline in mean brightness was observed after three of the four events
(the fourth occurred too shortly before the end of the observing
season).

0.0129\,c/d is not the only low frequency of 25\,Ori; it does not at
all times yield a good approximation of the SMEI light curve; the
events in 2016/17 are not clearly phased with it, and they look so
different that, without reconfirming time-series analysis, the light
curve could be from some other star.  Even the two events in 2014/15
differ by a factor of 2 in amplitude.  Therefore, 0.0129\,c/d is not
the only ruler of 25\,Ori's long-term variability.  Similarly to
0.1777\,c/d, 0.0129\,c/d is the difference frequency of several pairs
of strong peaks in group $g_1$.  Because of the very large amplitudes
of the events associated with the 0.0129\,c/d variability, some of the
weaker components of these pairs may be side lobes.  The symmetrical
location of such features on both sides of some strong peaks supports
this possibility whereas in some cases equal strengths of central and
neighboring peaks put it into question.  With amplitudes of 4.5 and
4.9\,mmag, respectively, the two components of the pair at 1.2963 and
1.3110\,c/d are among the three strongest peaks in group $g_1$ of
25\,Ori.

Because of the multitudes of shared difference frequencies, pairwise
frequency differences were calculated for numerous frequencies in both
$g_0$ and $g_1$.  $g_0$ was defined as 0-0.25\,c/d and $g_1$ was taken
as 1.1-1.75\,c/d.  Only differences falling into $g_0$ were included.
Matches were accepted as real if the deviation of a calculated
difference frequency from an observed frequency in $g_0$ was less than
10$^{-5}$\,c/d. In $g_0$ ($g_1$), there are 16 (12) frequencies with
individual amplitudes of $\geq$\,5\,mmag ($\geq$\,2\,mmag).  Thirteen
(twelve) of them form 24 (24) frequency pairs with others in the same
group such that for each pair the amplitude sum as well as the
amplitude associated with the difference frequency both exceed 10
(5)\,mmag.  Among the 16 frequencies in $g_0$ with amplitudes
$\geq$\,5\,mmag, 3 are such difference frequencies from $g_0$ itself
and 7 from $g_1$; the 3 frequencies are a full subset of the 7 and
have totals of six (0.0129\,c/d), nine (0.0122\,c/d), and eleven
(0.0074\,c/d) parents in $g_0$ and $g_1$.

The number as well as the fraction of frequencies involved in
combination frequencies grow very rapidly with both amplitude
threshold and matching tolerance in frequency.  The results above are
preliminary, and the analysis will be repeated with an entirely
independent algorithm and code (Baade et al., to be submitted to
A\&A).  The preliminary results also suggest that many frequencies in
$g_2$ (2.5-3.4\,c/d) are sums of frequencies in $g_1$; furthermore,
there are a few $g_1$ harmonics and higher-order harmonics from $g_0$.

\section{Discussion}
\label{discussion}
The BRITE and SMEI observations of 25\,Ori further refine the
conclusions derived earlier for $\mu$\,Cen, $\eta$\,Cen, and 28\,Cyg.
In the latter two stars, only two frequencies each were found to
combine to a high-amplitude difference frequency.  In 28\,Cyg, the
amplitude of the difference frequency, which was already high during
quiescence, was during the brightening multiplied within just 1-2
cycles.  In agreement with the conclusions for $\eta$\,Cen and
28\,Cyg, the elaborate pulsational model developed by
\citet{2017A&A...598A..74P} showed that the events seen in
KIC\,11971405 are not part of any beat pattern but result from genuine
(unexplained) amplitude amplification.  Peak-to-valley amplitudes of
100-200\,mmag during the events in 25\,Ori, which closely resemble
each other, provide further motivation to invoke mass loss as the
explanation.

However, photometry alone cannot diagnose genuine mass loss with any
certainty.  The nondetection during brightenings of phase shifts of
the difference frequency makes it impossible to separate immediate
pulsational effects on the photosphere and secondary effects due to
matter elevated to exophotospheric levels.  Neither can it be stated,
without detailed modeling, whether possible azimuthal temperature
variations in the photosphere with the difference frequency lead to a
similar modulation of the flux emitted by ejecta.  Perhaps, the slow
fadings after brightenings are the most robust indicator of matter
temporarily added to the inner disk and subsequently removed by
viscosity and/or radiation pressure.  All in all, it is very plausible
to think of brightenings as extra light re-emitted by ejecta
\citep{2012ApJ...756..156H} from mass-loss events.  Only parallel
high-cadence photometry and spectroscopy will lead to an unambiguous
conclusion.  However, the earlier spectroscopy of $\mu$\,Cen gave
already very clear hints in favor of this idea.

With the hindsight of the photometric difference frequencies, one
would perhaps describe the mass-loss activities of $\mu$\,Cen in terms
of two difference frequencies, 0.0180\,c/d and 0.0337\,c/d.  In
addition, variable line emission, indicating mass ejections, permitted
the inference of a threshold above which the superposition of NRP
velocity fields caused mass ejections \citep{1998ASPC..135..343R,
  2001A&A...369.1058R}.  This threshold of 15-20\,km/s for the
amplitude sum of the NRP modes involved was only exceeded for
combinations of the strongest with either the second- or the
third-strongest mode.  Superpositions of the second- and
third-strongest modes alone were ineffective.

Two more spectroscopic modes were found in $\mu$\,Cen
\citep{2001A&A...369.1058R}.  One of them had an amplitude that would
have been sufficient to let its combination with the strongest of the
first four modes exceed the mass-loss threshold.  However, these modes
seemed to be different ($\ell$\,=\,$m$\,=\,+3 vs.\
$\ell$\,=\,$m$\,=\,+2) so that the co-added vectorial velocity sum
reached its scalar value hardly anywhere.  It is only thanks to these
selection rules that the effects of combined NRP modes on mass loss
could be derived so clearly.  Otherwise, the star could have appeared
to be in permanent outburst as it does present itself in broad-band
photometry \citep{2016A&A...588A..56B} which is lacking an equivalent
of velocities as a means to distinguish ejecta from other matter in
the disk.  The numbers of frequencies visibly involved in the
mass-ejection process from $\mu$\,Cen and 25\,Ori are smaller than
those in $\eta$\,Cen and 28\,Cyg.  However, the greater simplicity of
the latter two stars may well be only apparent if the sensitivity of
the observations used for their analysis was not sufficient.

In 25\,Ori, the frequency seems to have been found of the clock that
opens and closes the mass-loss valve (\#\,3 in the list in
Sect.\,\ref{introduction}), namely 0.0129\,c/d.  Its value does not
appear to be numerically related to 0.1777\,c/d.  The significant
frequencies nearest to their linear combination frequencies,
0.1648\,c/d and 0.1806\,c/d, are 0.1643\,c/d (semi-amplitude
3.3\,mmag) and 0.1833\,c/d (5.4\,mmag).  Therefore, 0.1648\,c/d is a
marginal second-level difference frequency whereas the mismatch seems
too large for the possible sum frequency.

It may be of concern that the relevance of low difference-frequencies
in Be stars is discussed for BRITE/SMEI observations only.  However,
low-frequency variations are often removed up-front because they are a
priori suspect of being artefacts, or in power spectra they may appear
as unwanted side lobes of higher frequencies.  In other cases,
combination frequencies are just discarded because they are of no
obvious asteroseismic interest.  One exception is the recent
analysis of the {\it Kepler} observations of KIC\,11971405 by
\citet{2017A&A...598A..74P} who mention a frequency of 0.27644(2)\,c/d
which had one of the 4 largest amplitudes found in the full dataset.
The amplitude declined from 1800\,ppm to 900\,ppm after two events of
enhanced variability that P\'apics et al.\ call outbursts.  The middle
value between maximum and minimum read off Fig.\,22 of P\'apics et
al.\ is 0.2761\,c/d with a full range of $\pm$0.0004\,c/d.  These
authors' Table\,12 contains two series of frequencies.  Guided by the
BRITE/SMEI experience with other Be stars, the frequencies with the
largest amplitude were searched for the occurrence of 0.2761\,c/d as a
difference frequency.  The frequency with the second-largest amplitude
in the first group, 2.17192(2)\,c/d, and the highest-amplitude
frequency in the second group, 1.89575(2)\,c/d, differ by
0.27617\,c/d.  That is, most probably, this frequency is a difference
frequency.

0.276260\,c/d was found by \citet{2015MNRAS.450.3015K} and can be
described as the sum of two difference frequencies.  As in 28\,Cyg and
25\,Ori, this variability seems to be phase coherent (see Fig.\,26 in
P\'apics et al.) through the two events when its amplitude temporarily
increased.  The latter are reported as separated by 76.81\,d but
events are not repetitive on this timescale.  There were also various
smaller events which all occurred at different intervals.

Contrary to other pulsating stars, difference frequencies of Be stars
appear to be of twofold practical relevance: (i) They are involved in
the driving of mass loss.  (ii) Isolated frequencies without relation
to combination frequencies may be circumstellar frequencies as there
is no plausible way for extrastellar variabilities to be part of the
stellar grid of combination frequencies.  It is important to
distuinguish these two categories because they are diagnostics of
completely different physical processes.  In some Be stars, temporary
so-called {\v S}tefl frequencies \citep{1998ASPC..135..348S,
  2002A&A...388..899N, 2009A&A...506...95H, 2016A&A...588A..56B} seem
to trace the amount of matter in non-circularized and/or azimuthally
inhomogeneous near-stellar orbits \citep{2003A&A...401..271C}.  No
such frequency with amplitude $\geq$2\,mmag was identified in 25\,
Ori.

The apparent nature as combination (or harmonic) frequencies of many
high-amplitude frequencies in groups $g_0$ and $g_2$ may suffer from
chance coincidences of unresolved frequencies.  However, the bulk of the
identifications should be real since high-amplitude parent and
difference frequencies were selected for the analysis.  Frequency
groups also occur in SPB and $\gamma$ Dor stars.  They have been
attributed to rotation \citep{2011MNRAS.413.2403B} although most SPBs
do not rotate particularly rapidly
\citep[e.g.,][]{2017A&A...598A..74P} whereas very rapid rotation is a
strong requirement for the development of decretion disks in Be stars
\citep{2013A&ARv..21...69R}.  Another possible speculation is,
therefore, that high-amplitude combination frequencies play a role in
the selection of those modes in Be stars that are highly unstable
among a vast number of other modes with amplitudes not exceeding the
detection thresholds.

\section{Conclusions}
\label{conclusions}
Various studies have suggested that outbursts of Be stars may be
driven by many modes that are once in a while randomly constructively
cophased \citep[e.g.,][]{2015MNRAS.450.3015K}.  The converse causality
has also been proposed, namely that an outburst may lead to the
temporary enhancement or excitation of numerous pulsation modes
\citep{2009A&A...506...95H}.  The picture derived fom BRITE/SMEI
observations of $\eta$\,Cen, 28\,Cyg, and 25\,Ori and the spectroscopy
of $\mu$\,Cen is much simpler and does not depend on a strong random
component.  It is based on just a handful of modes which
hierarchically control the mass-loss process on four different
frequency scales $\nu$ that roughly differ by an order of magnitude:
\begin{list}{--}{\itemsep=0mm\parsep=0mm\topsep=0mm}
\item[1.)]  $\nu$\,$\sim$\,$10^{0}$\,c/d: many $g$ modes.  Example
  from 25\,Ori: 1.5014\,c/d.
\item[2.)]  $\nu$\,$\sim$\,$10^{-1}$\,c/d: pairwise difference
  frequencies of $g$ modes.  During outbursts, the amplitudes of a
  very small number of them may increase by up to an order
  of magnitude and contribute to the driving of the mass loss.
  Example from 25\,Ori: 0.1777\,c/d.
\item[3.)]  $\nu$\,$\sim$\,$10^{-2}$\,c/d: A very small fraction of
  these low difference-frequencies trigger the outbursts (open the
  hypothetical valve) by temporarily and quickly increasing the
  amplitude of the variations with the higher level-1 difference
  frequency.  Example from 25\,Ori: 0.0129\,c/d.
\item[4.)]  $\nu$\,$\sim$\,$10^{-3}$\,c/d:  On this timescale, the
  triggering of outbursts (opening of the valve) is cyclicly suspended
  and activated.  It could be another difference frequency, here
  between extremely closely spaced frequencies of individual NRP
  modes, or the presumably nonlinear combination of several level-2
  variabilities.  
\end{list}
Each of clocks 2-4 may have variable power, depending on whether or
not further NRP modes enhance or obstruct the main process.  These
additional modes may be drawn randomly from the general NRP floor or,
more likely, from the modes sharing the same difference frequency
(frequencies).  If the frequency network does not at all times behave
like a perfect spreadsheet, small frequency variations may result in
disproportionately large amplitude variations.  The presence of an
amplitude threshold for events as in $\mu$\,Cen could add a strong
nonlinear element.

The putative 4th level is lacking direct empirical support and only
based on an extrapolation of the first three levels.  An alternative
could be recovery phases in the driving of leaking NRP modes as
proposed by \citet{2014IAUS..301..173S}.  Clock levels 1-3 are in
agreement with the idea since the apparent pulsational driving of mass
loss implies considerable energy losses.  However, the postulated
temporary cessation of NRP variations seems unsupported by their
observed long-term presence \citep[e.g.,][]{2003A&A...411..167S,
  2016A&A...588A..56B}.

When the star-to-disk mass transfer is shut off (the valve opens too
rarely and/or too little), the disk quickly transforms itself from a
decretion to an accretion disk \citep{2012ApJ...744L..15C}.  Radiative
ablation is expected \citep{2016MNRAS.458.2323K} to further accelerate
the destruction of the disk.  The accompanying decrease, and
ultimately disappearance, of line emission would be the consequence.
In this way, disk life cycles of 10 or more years could be built from
elementary clocks running more than a thousand times faster.  Even
very simple and clean light curves \citep[][their
Fig.\,5]{2002AJ....124.2039K} demonstrate that several such clocks may
act in parallel, leading to the coexistence of several long timescales
ruling the emission-line strength.  If there are also multiple level 2
and 3 clocks, the behavior of a Be star may soon look erratic.

The apparent elimination as stellar eigenfrquencies of many
frequencies in $g_0$ and $g_2$ and the implied strong reduction of the
overall variability spectrum mostly to $g$ modes in $g_1$ indicates
another large simplification of the description that needs to / can be
given of the once enigmatic variability of some Be stars, one and a
half centuries after their discovery by \citet{1866AN.....68...63S}.

Many questions remain though.  Currently, most pressing among them
are:
\begin{list}{--}{\itemsep=0mm\parsep=0mm\topsep=0mm}
\item How broadly representative is the above simplified picture of Be
  stars? 
\item Is it applicable to late-type Be stars which in ground-based
  observations appear much less active?  The case of the B8e star
  KIC\,11971405, which is the only late-type Be star studied at this
  level of detail, suggests that at least some late-type Be stars do
  agree with it.
\item
Does the distribution of NRP frequencies select the difference frequencies 
involved in the mass-loss process, or do atmospheric timescales favoring 
large-amplitude variations filter the NRP spectrum?  
\item How do amplitudes of difference frequencies grow during
  brightenings by up to an order of magnitude and 3-5 times more
  strongly than those of the parent frequencies?
\item What is the stellar/circumstellar contribution to this amplitude
  amplification?
\item
  What is the nature of the postulated valve?  Is it a threshold in the
  combined amplitude of the NRP modes involved? 
\end{list}

\bibliographystyle{ptapap}
\bibliography{baade}

\end{document}